\documentclass[12pt]{article}
\usepackage[english]{babel}
\usepackage{amsmath}
\usepackage{times}
\usepackage[body={6.5in,9.5in}]{geometry}

\usepackage{amsmath, amsthm, amssymb,float}
\usepackage{amsfonts}
\usepackage{graphicx}
\usepackage{dcolumn}
\usepackage{bm}
\usepackage{url}
\usepackage{textcomp}
\usepackage[normalem]{ulem}
\usepackage{mathrsfs}
\usepackage{url}
\usepackage{amsmath}
\usepackage{graphicx}
\usepackage{subfigure}
\usepackage{color}

\title{\it Notes on a  1-dimensional electrostatic  plasma model}
\author{{F. Pegoraro$^1$, P.J. Morrison $^2$}\\
{\small  $^1$Department of Physics, University of Pisa, Pisa, Italy}\\
{\small $^2$Physics Department,          University  of Texas at   Austin, Austin, TX}}
\date{}

\begin{document}
\maketitle

\begin{abstract}
 A starting point for deriving the Vlasov equation  is the BBGKY 
hierarchy that describes the dynamics of coupled marginal distribution functions. \, With a large value of the 
plasma parameter one can  justify eliminating  2-point correlations in terms of the 1-point 
function in order to derive  the Vlasov Landau Lenard  Balescu (VLLB) 
theory. 
 Because of the high dimensionality of the problem, numerically testing the assumptions 
of the VLLB theory is prohibitive. \,
In these notes  we propose  a physically reasonable 
interaction model  that  lowers the dimensionality of the problem  and   may bring such computations within reach.
We  introduce a 1-dimensional (1-D)  electrostatic  plasma model formulated in terms of the interaction of parallelly-aligned charged disks.  This model combines  1-dimensional features at short distances and 3-dimensional features at large distances. 

\end{abstract}

\section{Introduction}\label{intr}
The construction  of a charged-disks  electrostatic model is motivated  by  the aim \cite{wtw} to  verify numerically the 
 validity  of the procedure which is adopted, see e.g. Ref.\cite{Krall},  in order to solve  the BBGKY hierarchy \cite{BBGKY}  in a weakly coupled plasma  when  deriving the kinetic plasma equation, i.e. the Vlasov equation.    Such a verification might be  based on   the numerical  integration of the time evolution of the two-point particle distribution function  as obtained from  a proper Hamiltonian truncation of the the BBGKY hierarchy  at the level of the three-point distribution function \cite{Morrison}.  In 3-D space the equation for  the  time evolution of this two-point   distribution  function
 would be 13-dimensional (one time  dimension plus  the 12-dimensional two-particle phase space).  In 1-D space the  equation for  the two-point  (actually two foil) distribution   would be 5-dimensional (time plus 4-dimensional two-foil
phase space).  However in 1-D the electric field generated by a charge foil does not decay with distance so that the  two-foil interaction energy diverges at infinity. 
This makes it  impossible  to define  a finite correlation length. 
A possible compromise is to consider the interaction of  aligned, uniformly charged disks, a sort of a  single-row abacus made of  thin disk-shaped beads, where we can introduce a characteristic length $b$, the radius of the disk, which we may identify with the  Debye radius $\lambda_{d}$ (same for the electrons and ions).  In this case the electrostatic interaction energy  between two charged disks is finite both at zero and at infinite distance $\xi$  from the source and its absolute value is a decreasing function of $\xi$.

The present notes are  dedicated to the derivation of the main properties of such a disk-plasma in order to clarify where and how the dynamics of such a model system  can mimic  the dynamics of a real plasma, at least as long as the longitudinal electric field limit is concerned. In analogy with a real  particle plasma,  in the  derivation which follows we will  introduce concepts such as positive and negative charged disk densities,  collective electric field shielding, disk waves  etc., but  we will not use a Poisson type equation with the  charge disk density  as the source. It will be enough to refer  to the expression of the interaction energy between the disks derived in Ref.\cite{Disk} that  in the present model will  play the role of the inter-particle Coulomb energy in a real 3-D plasma.

\section{Disk  Model}\label{potential}

\begin{figure} [h!]\center
\includegraphics[width=0.7\textwidth]{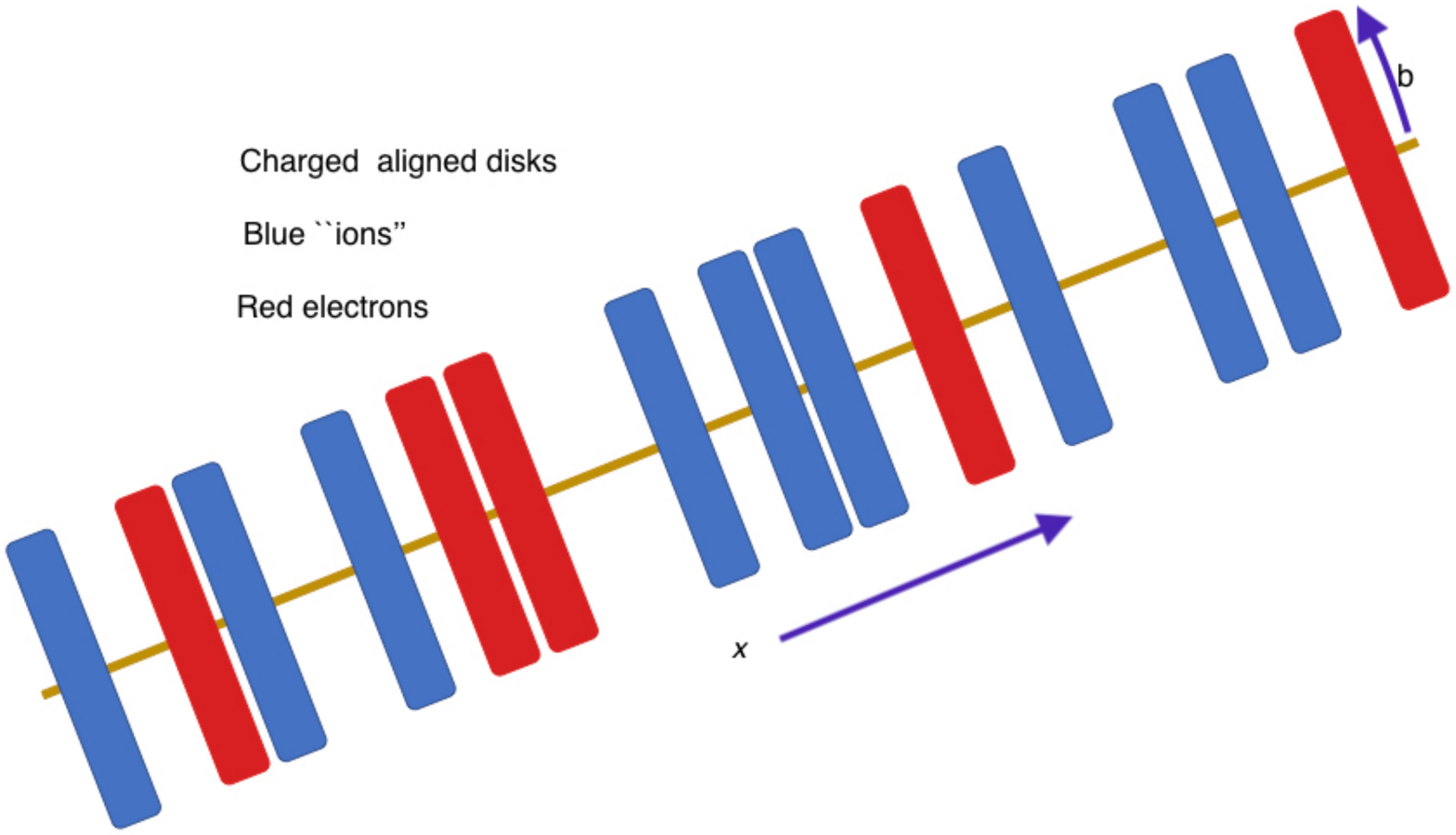}
\caption{Schematic view, not in scale, of the disk configuration with the different colours representing different charges.}
\label{CP0}
\end{figure}

The ``disk model''  describes  a globally neutral system consisting of a large number (as calculated  inside intervals  with length  of the order  of the disk radius $ b$) of   aligned,  infinitely thin disks with equal  radius,  as sketched in Fig.\ref{CP0}, equal or opposite  electric charge and different masses (ion-disks, electron-disks) that are free to move along the $x$ axis (the planes of the disks are orthogonal to the $x$ axis) and to  pass through each other. 

In the following, for the sake of simplicity, the  ion-disks are taken to be immobile.
We denote the mass of an electron disk by $M_e$ and its charge by $Q_e =- Q_i$ with $ Q_i$ the charge of an ion disk.

\subsection{Discrete disks: interaction energy}

The interaction energy ${\cal W} (x_1,x_2)$  between two uniformly  charged  infinitely thin disks  of radius $b$ located  at $x_1$ and $x_2$ respectively can be written in c.g.s.  units as  \cite{Disk}. \begin{equation}\label{1}
{\cal W } (x_1,x_2) =  4  \frac{Q_1\,  Q_2}{b} V(\xi) = Q_1 \varphi_2 =  Q_2 \varphi_1,
\end{equation}
where
\begin{equation} \label{2}
V(\xi) = - \frac{\xi}{2} \left\{  1 - \frac{1}{3\pi} \left[ (4 -\xi^2) E(-4/\xi^2)  +  (4 +\xi^2) K(-4/\xi^2) \right]   \right   \}.
\end{equation}
and $\varphi_1$, $\varphi_2$ are the potentials generated by the disks $1$ and $2$ respectively.
Here $\xi(x_1,x_2) = |x_1 - x_2|/b$,  $ Q_1=\pi b^2\sigma_1$ and $ Q_2 = \pi b^2\sigma_2$ are the (fixed) charges  of the two disks, $\sigma_{1,2}$ are their surface density and   $ E(-4/\xi^2) $  and $ K(-4/\xi^2) $  are elliptic integrals \cite{abram}. The ``disks-electrostatics'' is illustrated  in more detail in Appendix \ref{App1}.
A plot of $V(\xi)$ is given in Fig.\ref{figV} .
\begin{figure} [h!]\center
\includegraphics[width=0.4\textwidth]{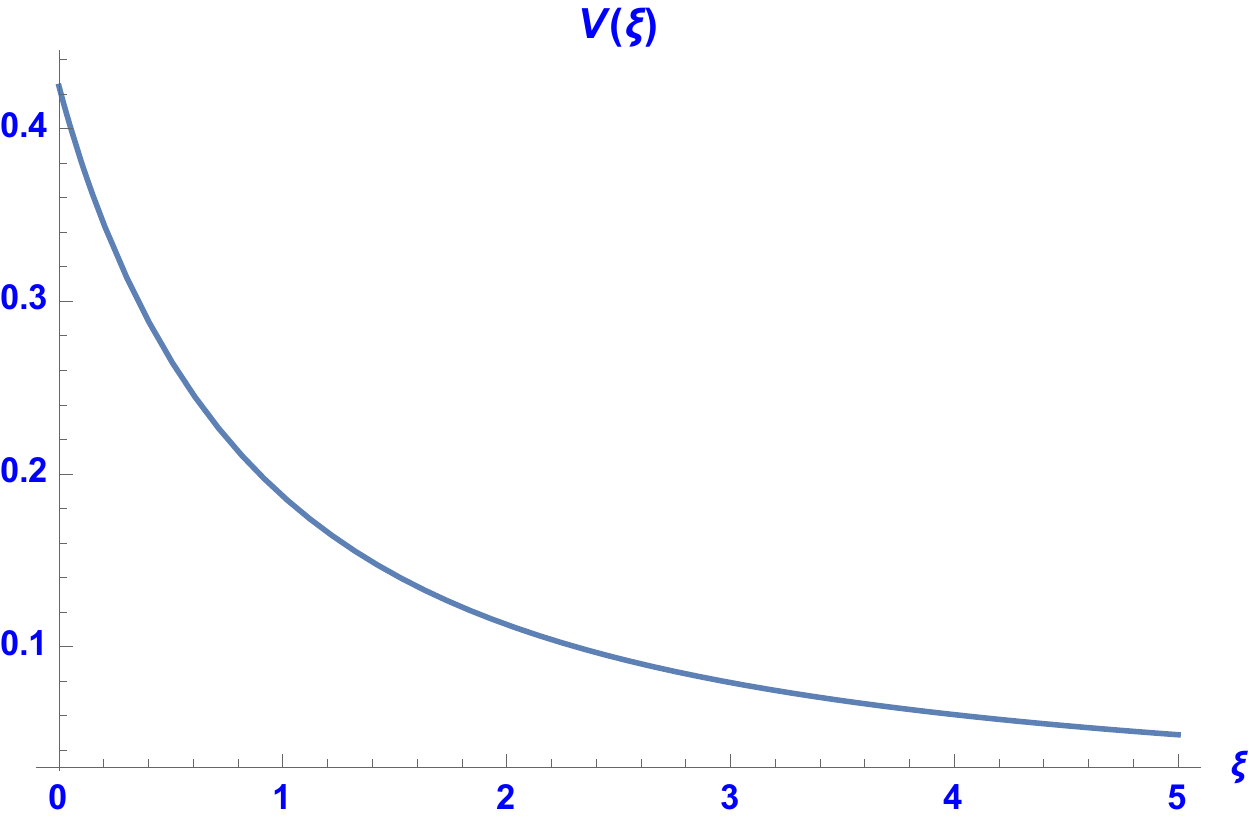}
\caption{Plot of the potential $V(\xi) $ versus $\xi$.\,   Note the $1/(4\xi)$-behavior for large $\xi$ and the finite value $4/(3 \pi)$ at $\xi = 0$.}
\label{figV}
\end{figure}

\begin{figure} [h!] \center 
\includegraphics[width=0.4\textwidth]{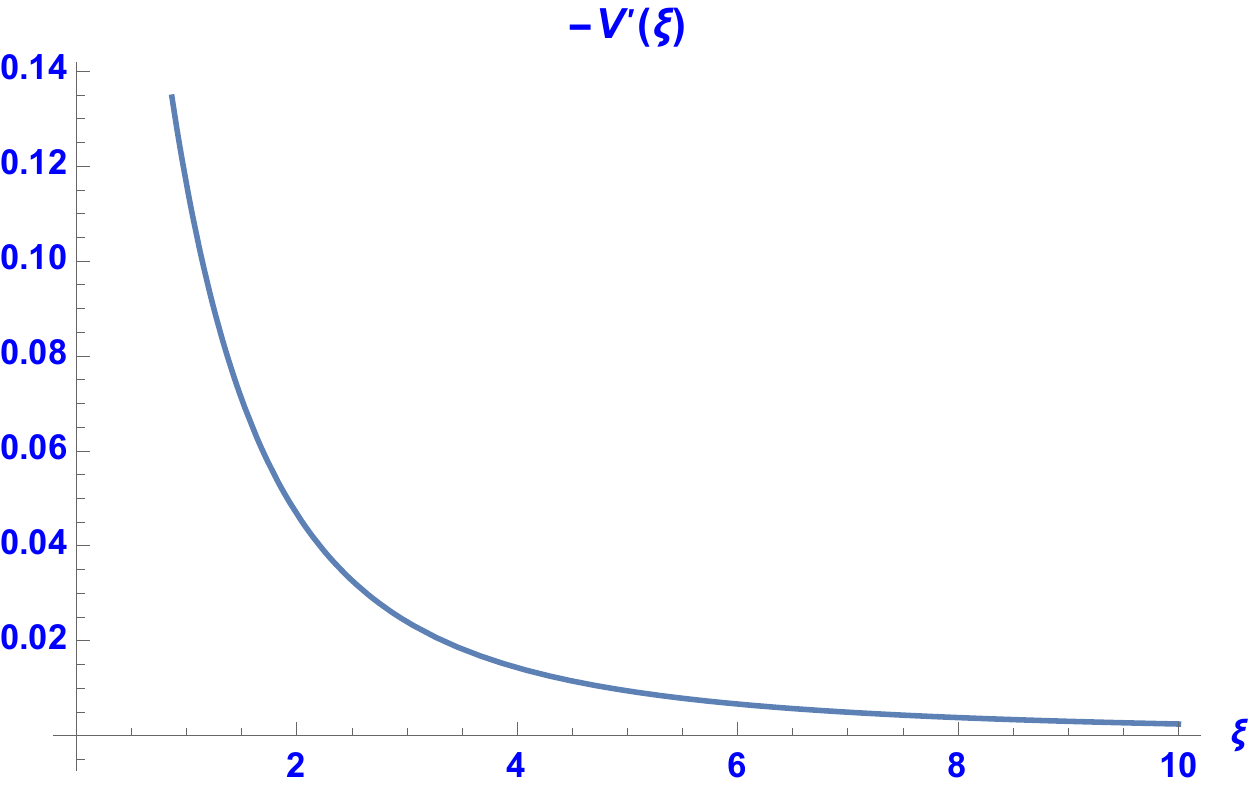}
\includegraphics[width=0.4\textwidth]{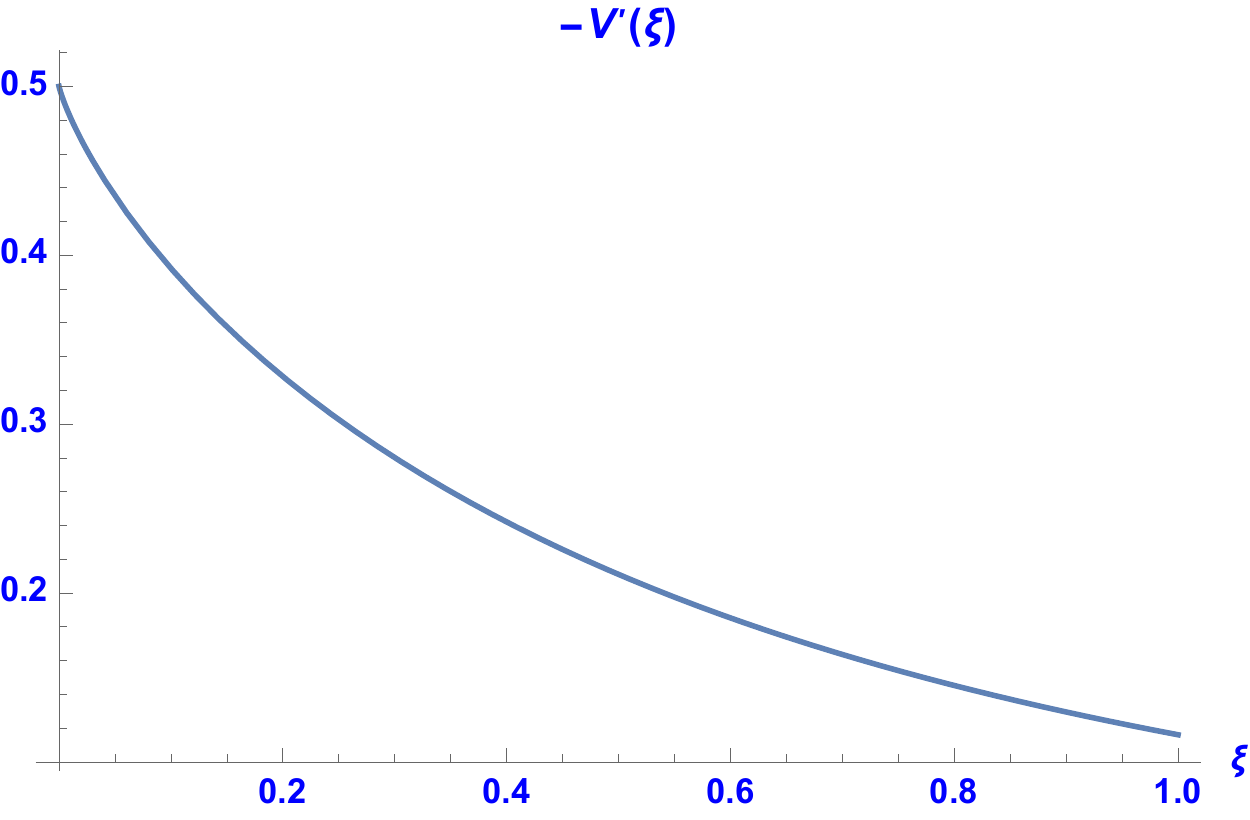}
\caption{Plot of  $-V'(\xi) $ versus $\xi$.\,   Note the $1/(4\xi^2)$ behavior for large $\xi$ (left frame) and the value $-V'(0) = 1/2$  (right frame).}
\label{figV'}
\end{figure}

The corresponding electric force ${\cal F}_{Q_2}$ acting on the disk 1  due to the disk 2 is given by 
\begin{align} \label{3}&  {\cal F}_{Q_2} (x_1) = -  {\cal F}_{Q_1} (x_2)=  -(4  Q_1 Q_2/b^2) \,  V'(\xi) \, {\rm sign} \,(x_1 -x_2) ,\qquad  {\rm where} \quad \\ &
V'(\xi) =\frac{d V(\xi)}{ d\xi }=  \left[\frac{1}{2}  +  \frac{1}{2\pi} \left[ \xi^2 E(-4/\xi^2)  -  (4 +\xi^2) K(-4/\xi^2) \right]   \right]. \nonumber \end{align}
A plot of $-V'(\xi)$ is given in Fig.\ref{figV'}

\section{Continuous limit}

Define an electron-disk density $n_e(x,t)$  (number of disks per unit length) and an ion-disk density $n_i(x,t) = n_i$ (for simplicity stationary and homogeneous).
This defines a disk-charge density (charge per unit length)
\begin{equation} \label{4}
\rho(x) = Q (n_i -  n_e(x,t)),
\end{equation}
with $Q = Q_i= -Q_e$.

Define the electron-disk velocity $u_e(x,t)$ that satisfies the continuity equation
\begin{equation} \label{5}
\partial_t n_e(x,t) + \partial_x [n_e(x,t) \, u_e(x,t) ] = 0.
\end{equation}

Assuming that $n_i -  n_e(x,t)$ vanishes (sufficiently fast) at $\pm \infty$,  the  electrostatic potential  created by a  disk-charge density  can be written in convolution integral form as 

\begin{equation} \label{6}
\varphi(x)  =  (4/b)  \int_{-\infty}^{+\infty} dx' \rho(x')V(|x-x'|/b),
\end{equation}
which returns  for $\rho(x')  = Q \delta (x' -x_1)$ the expression for the potential of a disk given in  Eq.(\ref{1}).

The corresponding  electric force acting on a disk of charge $Q_j$ \, ($j= i,e$)   can be written as 
\begin{equation} \label{7}
{\cal F}_j(x)  =  -(4 Q_j/b) \frac {d}{dx} \int_{-\infty}^{+\infty} dx' \, \rho(x')V(|x-x'|/b),
\end{equation}
which returns  Eq.(\ref{3})  for $\rho(x')  = Q_1 \delta (x' -x_1)$.
To close the system we  may supplement the continuity equation (\ref{5})  with the   momentum equation of a ``cold''  disk plasma
\begin{equation} \label{14a}
M_e n_e\left( \partial_t + u_e \partial_x\right) u_e =  n_e {\cal F}_e.
\end{equation}

In  the rest of this section we find it convenient to use  dimensionless dynamical quantities based on a characteristic  time (the inverse of the  disk  plasma frequency\footnote{\label{N1} This expression returns the standard definition of the plasma frequency after the following identifications:  $Q/M_e =  e/m_e$ and $Q \, n_o /(\pi b^2) = \sigma \, n_o = e \, n_{vol}$ where $\sigma = Q  /(\pi b^2)$ is the disk surface charge  and  $ e \,  n_{vol} $ is the volume charge obtained by multiplying the surface charge  of  the disk times the number of disks  per unit length $n_o$.}, see Eq.(\ref{16})) 
\begin{equation} \label{5dim}\omega_{Dpe}^{-1}= \left[ {4 Q^2 n_o}/{M_e b^2}\right]^{-1/2} ,
\end{equation}
where $n_0$ is the (uniform) background density,  and on the length $d$ (note that the disk electrostatic, contrary to standard electrostatic,  has a characteristic spatial scale).\, \, 
Specifically we set
\begin{equation} \label{5dim1}
\tau =  t \, \omega_{Dpe}, \quad {\rm and }\quad  X =  x/b,
\end{equation}
in agreement with the definition of the variable $\xi$.
Then we define 
\begin{align} \label{5dim2} &
N_{e,i}(X,\tau) = b  \, n_{e,i}(x,t), \quad \varrho(X,\tau) =  b\,  \rho(x,t) = Q (N_i - N_e), \ \\&
u_{e}(x,t)  =    v_D  \,  U_e (X,\tau)  \quad  {\rm where} \quad   v_D = b \, \omega_{Dpe}= [{4 Q^2 n_o}/{M_e}]^{1/2},  \quad  \Phi(X) =  b\, \varphi(x)/4, \nonumber \\&
 \quad   F_e = b^2/ (4 Q^2 N_0) {\cal F}_e(x)  =  \frac{1}{N_0 Q} \frac{d}{dX} \int_{-\infty}^{+\infty} dX' \, \, \varrho(X')V(|X-X'|/b), \nonumber \end{align}
where $\varrho(X')/ (N_0 Q)= (N_i - N_e)/N_0$ is dimensionless. Note that in a  particle plasma there is no velocity directly corresponding to $v_D$. 
Then Eqs.(\ref{5},\ref{14a}) read
\begin{align} \label{5bis} &
\partial_\tau N_e(X,\tau) , + \partial_X [N_e(X,\tau) \, U_e(X,\tau)  ] = 0, \nonumber \\&
N_e\left( \partial_\tau + U_e \partial_X\right) U_e =  F_e,  \end{align}
with \begin{equation} \label{equations}
F_e =   \frac{d}{dX} \ \int_{-\infty}^{+\infty} dX' \,\frac{N_i - N_e(X')}{N_0 }\, V(|X-X'|).  \end{equation}

\subsubsection{Laplace transform representation}

Start from Eq.(\ref{6}) with  the potential 
$\varphi(x)  =  (4/b)  \int_{-\infty}^{+\infty} dx' \rho(x')V(|x-x'|/b)$
and write  the  integral  in dimensionless form
\begin{equation}\label{Lap1}
\int_{-\infty}^{+\infty} dx' \rho(x')V(|x-x'|) =  \int_{0}^{+\infty} ds  [J_1(s)/s]^2 \, \int_{-\infty}^{+\infty} dx' \rho(x')\exp{(-s(|x-x'|)} ,
\end{equation}
where  $J_1$ is a Bessel function and we used Eq.(\ref{A4}). The double-sided Laplace transform in $dx'$ can be written as 
\begin{align}\label{Lap2}
&\int_{-\infty}^{+\infty} dx' \rho(x')\exp{(-s(|x-x'|)} = \int_{x}^{+\infty} dx' \rho(x')\exp{[s(x-x')]}  + \int^{x}_{-\infty} dx' \rho(x') \exp{[s(x'-x)]} \nonumber\\
& = \int_{0}^{+\infty} dy\, \rho(x+y)\exp{[-sy]} +  \int_{-\infty}^{0} d\eta\, \rho(x+\eta)\exp{[s\eta]} \nonumber \\&
 =  \int_{0}^{+\infty} dy\,[ \rho(x+y) +\rho(x-y)] \exp{[-sy]}.\end{align} 
Thus we obtain 
\begin{equation}\label{Lap3}
\int_{-\infty}^{+\infty} dx' \rho(x')V(|x-x'|) =  \int_{0}^{+\infty} ds  [J_1(s)/s]^2 \, \int_{0}^{+\infty} dy\,[ \rho(x+y) +\rho(x-y)] \exp{[-sy]}.
\end{equation}
Computing the integral over $s$ first would bring us essentially to the initial Eq.(\ref{6}).  On  the contrary it is convenient to use the fact 
that  Eq.(\ref{Lap3}) contains a special combination of Poisson transforms of the density $\rho(x)$ and expand   $\rho(x)$ on a basis with elements that transform in a simple way under Poisson  transform.

Thus we write the Fourier series (finite domain, periodic boundary conditions)
\begin{equation}\label{Lap4}
\rho(x) = \Sigma_{n=0}^\infty  [C_n\cos{ (n x) } + S_n\sin{ (n x) } ],
\end{equation}
 use the relationships 
 \begin{align}\label{Lap5}
& \int_{0}^{+\infty} dy\, \cos{(ay +b)} \exp{[-sy]} =\frac{s \cos{(b)} - a\sin{(b)}}{s^2 + a^2} \nonumber\\
& \int_{0}^{+\infty} dy\, \sin{(ay +b)} \exp{[-sy]} =\frac{s \sin{(b)} + a\cos{(b)}}{s^2 + a^2}.
\end{align}
and obtain 
\begin{align}\label{Lap6}
& \int_{0}^{+\infty} dy\,[ \rho(x+y) +\rho(x-y)] \exp{[-sy] }=\\
&\int_{0}^{+\infty} dy  \exp{[-sy]}\, \ \Sigma_{n=0}^\infty \{C_n[\cos{ [n( y +  x)]  +\cos[n(  y -x)} + S_n[\sin{ [n(y + x) ]} -\sin{[n (y -x)] } \}
\nonumber\\
&  =   \Sigma_{n=0}^\infty \,  \frac{2s}{s^2 + n^2} [ C_n \cos{(nx)} + S_n \sin{(nx)}]. \end{align}
Thus finally we have
\begin{equation}\label{Lap7}
\int_{-\infty}^{+\infty} dx' \rho(x')V(|x-x'|) =    \Sigma_{n=0}^\infty \,   \, [C_n \cos{(nx)} + S_n \sin{(nx)}] \, \int_{0}^{+\infty} ds   \frac{2 J^2_1(s)}{s(s^2 + n^2)},
\end{equation}
where the last integral is a known function (see Eq.(\ref{A5}.)

\subsubsection{Fourier space}

Assuming the required boundary conditions at infinity,  we write
\begin{equation} \label{8}
  {\hat \rho}(k) = \frac{1}{(2\pi)^{1/2}} \int_{-\infty}^{+\infty} dx \,  \rho(x) \, \exp{(-ikx)} ,\quad  {\hat{\cal F}}_j(k)= \frac{1}{(2\pi)^{1/2}} \int_{-\infty}^{+\infty} dx \,   {\cal F}_j(x)  \exp{(-ikx)}
\end{equation} 
 and obtain from Eq.(\ref{7}), which has the form of a convolution integral,
  \begin{align} \label{9} &
{\hat{\cal F}}_j(k )  = - \frac{Q_j}{4b}  \frac{1}{(2\pi)^{1/2}} \int_{-\infty}^{+\infty} dx \, \exp{(-ikx)} \frac{d}{dx}  \int_{-\infty}^{+\infty} dx' \, \rho(x')V(|x-x'|/b), \\&  =- \frac{4 Q_j}{b}  \frac{ik}{(2\pi)^{1/2}} \int_{-\infty}^{+\infty} dx \, \exp{(-ikx)}  \int_{-\infty}^{+\infty} dx' \, \rho(x')V(|x-x'|/b),\nonumber  =
- 4 i k  Q_j   {\hat \rho}(k)  {\hat  V}(kb), \nonumber
\end{align}
where  ${\hat V}(kb$) is defined (without including the factor $1/(2\pi)^{1/2}$) by 
\begin{equation} \label{10} 
 {\hat V} (kb) =  \int _{-\infty}^{+\infty} \frac{dx}{b} \,\, V(|x|/b) \exp{-[i (kb) \, (x/b)]}] =  2 \int _{0}^{+\infty} {dx}/{b} \, \, V(|x|/b) \cos{ [(kb) \, (x/b)}]
 \end{equation} 
 and is explicitly calculated in Appendix \ref{App1}. 
\begin{figure} [h!] \center 
\includegraphics[width=0.4\textwidth]{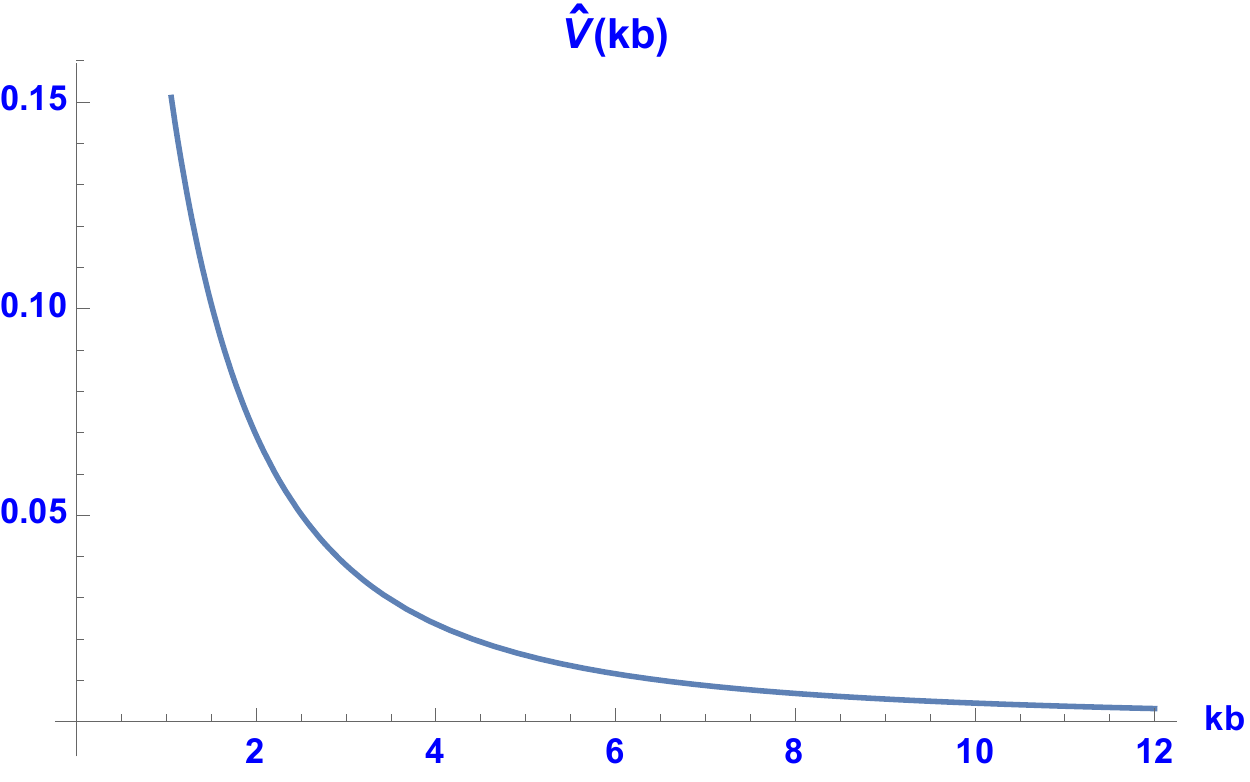},
\includegraphics[width=0.4\textwidth]{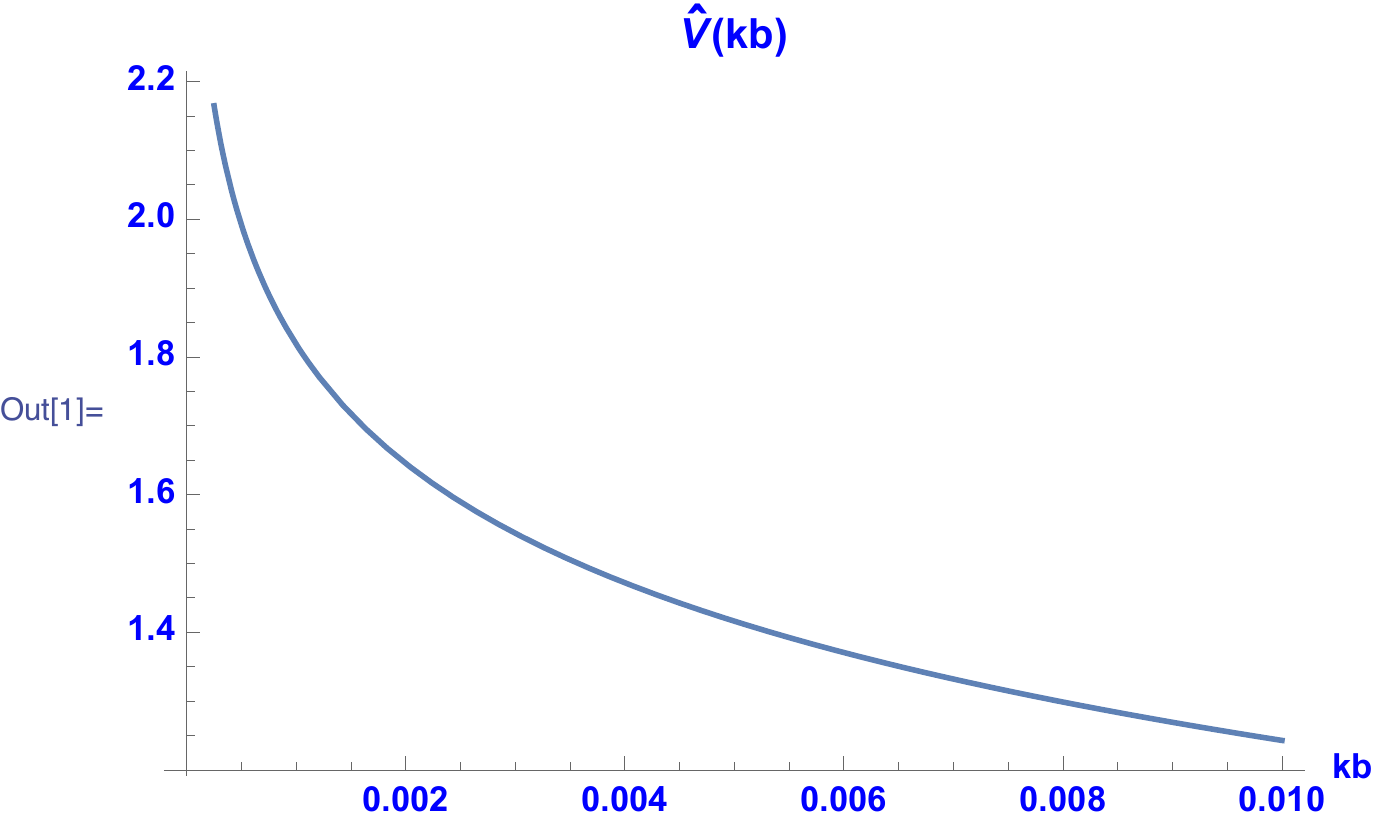}
\caption{Behavior of ${\hat V}(kb)$ for large (left frame) a  for small values (right frame) displaying the logarithmic contribution, see Eq.(\ref{A6}). }.
\label{G-small}
\end{figure}

\subsection{Screening at thermodynamic equilibrium}

In order to  determine the screening of a test disk of charge $Q_{t}$  at $x=0$ produced  by the spatial rearranging of the other disks  we consider 
 an electron-disk  Boltzmann equilibrium\footnote{In this case it is not correct to take the ions as immobile, but here for the the sake of illustration,  a factor of 2  may be disregarded.} expressed in terms of the screened potential $\varphi_{s}$
 \begin{equation} \label{11}
n_e(x) =   n_e(o) \exp{ [(Q \varphi_{s} )/ T_e ] } \sim  n_e(o) [ 1+Q \varphi_{s} / T_e ].
\end{equation}
Note that in this disk model, contrary to a Coulomb plasma,  the approximation $    Q \varphi_{s} / T_e \ll 1$ remains valid even at close distance. Setting $ n_i =   n_e(o) = n_o$
we can write the charge density of the  test disk  plus the screening density  obtained  from Eq.(\ref{11})  
 \begin{equation} \label{12}
\rho (x) =   Q_t\delta(x) - Q^2   n_o \varphi_{s} / T_e, 
\end{equation}
which inserted into Eq.(\ref{6}) gives the integral equation
\begin{equation} \label{13}
\varphi_{s}(x) =   Q_t  (4/b) V(|x|/b) -  4(Q^2 n_o/T_e)  \int_{-\infty}^{+\infty}  \frac{ dx'}{b } \varphi_{s}(x')V(|x-x'|/b).
\end{equation}
Since the integral term  in Eq.(\ref{13}) is a convolution product, Eq.(\ref{13}) can be solved by performing a Fourier transform with respect to $x$ (see Appendix \ref{App1} for analytical details of the functions involved).
Note that the factor in front of the integral is dimensionless  and can be interpreted as $\sim b^2/\lambda_d^2 \sim 1$ if we recall that $n_o$ is the number of disks per unit length and that the disk radius 
is $b\sim \lambda_d$.

\begin{figure} [h!]
 \centerline{\includegraphics[width=0.3\columnwidth]{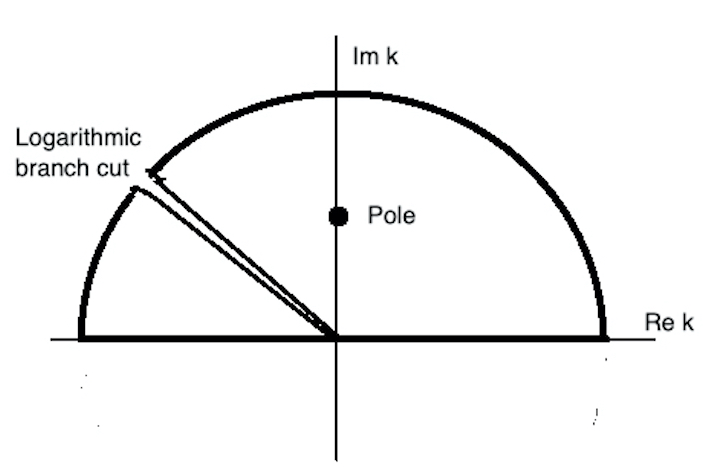}}
\caption{ Sketch of the Integration contour for the inverse Fourier transform.  of Eq.(\ref{FF13b})}
\label{fig1}
\end{figure}

The Fourier transform of Eq.(\ref{13}) reads
\begin{equation} \label{F13}
{\hat \varphi}_{s}(kb) =   Q_t  (4/b) {\hat V}(kb) -  4  (Q^2 n_o/T_e) {\hat \varphi}_{s}(kb)\,  {\hat V} (kb),  \end{equation}
where
\begin{align} \label{FF13}
& {\hat \varphi}_{s}(kb) = \frac{1}{(2\pi)^{1/2}} \int_{-\infty}^{+\infty} dx/b \,\, \varphi_s(x) \, \exp{-i[(kb)\, (x/b)]} , 
\\&   {\hat V} (kb) =  \int _{-\infty}^{+\infty} {dx}/{b} \,\, V(|x|/b) \exp{-[i (kb) \, (x/b)]}] =  2 \int _{0}^{+\infty} {dx}/{b} \, \, V(|x|/b) \cos{ [(kb) \, (x/b)}].  \nonumber
\end{align} 
The definition of ${\hat V} (kb)$ corresponds to $(2\pi)^{1/2}$ times the Fourier transform of $ V(X)$. From Eq.(\ref{F13}) we obtain
\begin{align} \label{FF13a}
&[ 1 +    4 (Q^2 n_o/T_e) \,  {\hat V} (kb)]  {\hat \varphi}_{s}(kb) =    Q_t  (4/b) {\hat V}(kb), \quad {\rm i.e.}  \nonumber
\\&  \varphi_{s}(x) =  \frac{1}{(2\pi)^{1/2}}\int _{-\infty}^{+\infty} {d(kb}) \,\frac{Q_t  (4/b) {\hat V}(kb)\exp{[i (kb) \, (x/b)]}}{1 +    4 (Q^2 n_o/T_e) \,  {\hat V} (kb)} \nonumber
\\&  
 =  \frac{2}{(2\pi)^{1/2}}\int _{0}^{+\infty} {d(kb}) \,\frac{Q_t  (4/b) {\hat V}(kb)\, \cos{[(kb)\, (x/b)]}}{1 +    4 (Q^2 n_o/T_e) \,  {\hat V} (kb)}.
\end{align} 
Defining the dimensionless temperature\footnote{This is the only dimensionless parameter of the theory and is equal to one if we model  a pressure isotropic plasma and take $b = \lambda_{d}$.}
$
\Theta =  [4  (Q^2 n_o/T_e)]^{-1}  = (v_{the_{x}}/v_D)^2 ,$
Eq.(\ref{FF13a}) can be written in dimensionless form as 
\begin{equation} \label{FF13b}
 \Phi_{s}(X)  =  \frac{2 Q_t }{(2\pi)^{1/2}}\int _{0}^{+\infty} {d\kappa} \, \frac{{\hat V}(\kappa)}{1 +   {\hat V} (\kappa)/\Theta} \, \,  \cos{(\kappa X)} .
\end{equation} 
It is not easy to compute the inverse Fourier transform (\ref{FF13b}) in explicit analytic form.   Taking into account that  ${\hat V}(kb)\propto 1/(kb)^2$   for large $(kb)$ and  if no additional  singularities  in ${\hat V}(kb)$ are present\footnote{G-Meijer functions  are  analytic aside possibly for $z = 0$, $|z|= 1$ and $z=\infty$, see Ref.\cite{Nist}} aside for a logarithmic term starting from the origin (see Eq.(\ref{A6})), the standard procedure, would   compute  the integral in Eq.(\ref{FF13b}) by using  a   Cauchy contour of the type sketched in in Fig.(\ref{fig1}). 

\bigskip

The residue at the  complex pole 
${\hat V} (kb)/\Theta = -1  $
would lead to an exponential shielding  analogous to the Debye shielding for a particle plasma.
 The integral along  the two sides of the logarithmic branch-cut would  lead to a non exponential contribution not present in a particle plasma.

\begin{figure}[h!]
\center 
{\includegraphics[width=0.3\columnwidth]{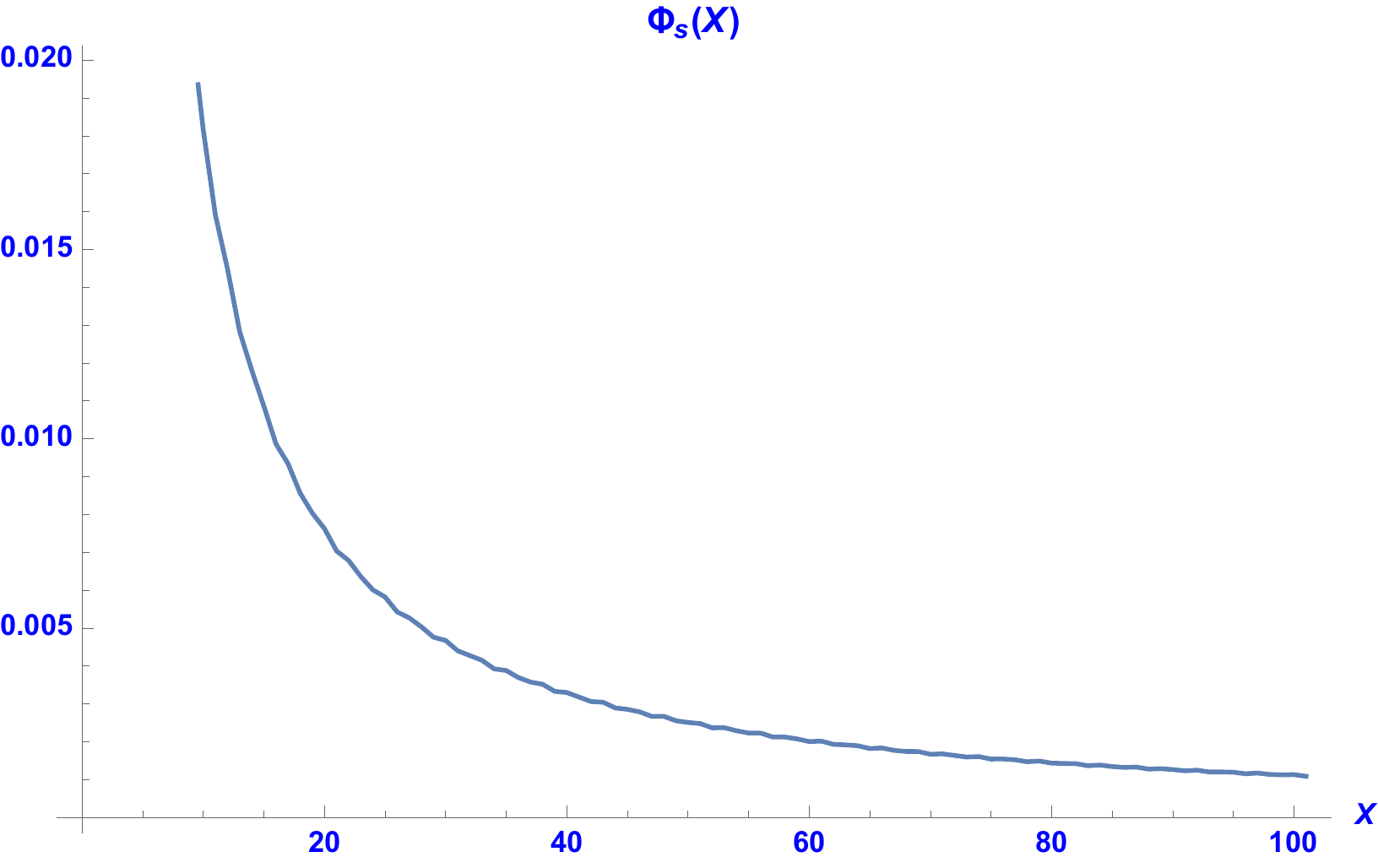}}
{\includegraphics[width=0.3\columnwidth]{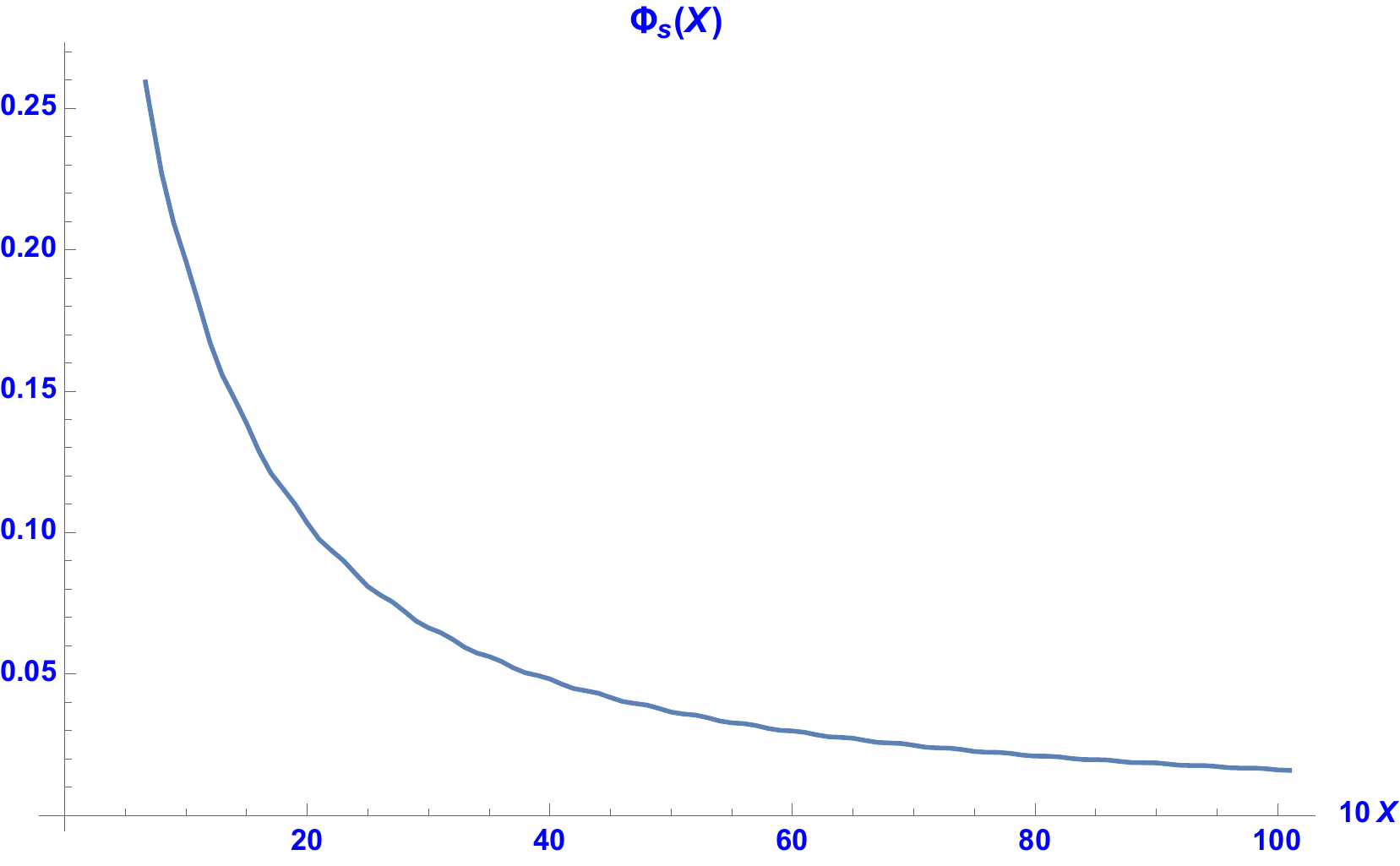}}
{\includegraphics[width=0.3\columnwidth]{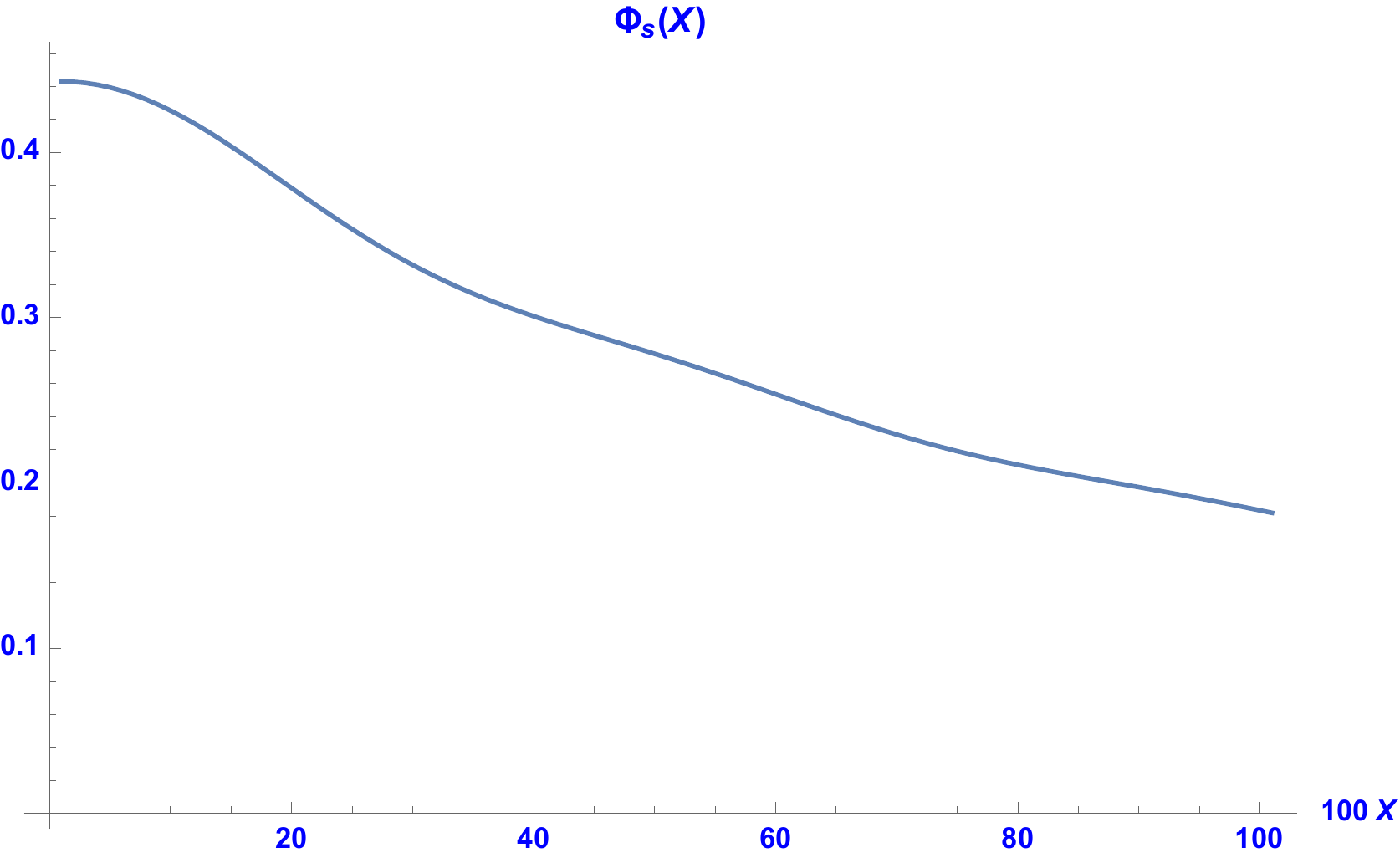}}
\caption{Plot of  the  screened  potential   $\Phi_{s}(X)$  in Eq.(\ref{FF13b}) as a function of $X= x/d$: left frame in the interval $0 - X - 100$, center frame enlargement in the in the interval $0 - X- 10$ and 
right frame in the interval $0 -X-1$. 
The integration over  $\kappa$ has been  limited to the interval, $0$-$15$ while the label  $100$ on the horizontal axis refers to the number of points in $X$  for which the  numerical integration has been performed.
}
\label{HW}
\end{figure}

 These qualitative statements  are confirmed by  integrating Eq.(\ref{FF13b}) numerically for different values of 
the dimensionless  temperature $\Theta$. In Fig.\ref{HW} we  show the result of this integration for $\Theta = 2$.
In agreement with  the  potential ${V}(\xi)$ in Eq.(\ref{2}), the shielded potential   $\Phi_{s}(X)$ is finite and nearly constant at $X\sim 0$ as shown in the enlarged plot in the right frame. For larger values of $X$ the potential $\Phi_{s}(X)$ has a fast  exponential-type decay.
 Afterwards,  as already visible in the central frame,  the spatial dependence of the screened potential
 changes  to a slower power-like decay, as shown by  the full plot in the left frame.  The exponent of this power law decrease is marginally larger in absolute value than $1$, i.e. marginally faster  than the non-shielded potential.

\section{Cold disk plasma  waves}

We consider the case where the disk thermal motion can be neglected and derive the linear dispersion relation of the cold disk plasma waves  using Eqs.(\ref{5},\ref{6},\ref{7},\ref{14a}).

Linearization for a monochromatic  wave in a stationary equilibrium with density $n_0$  and immobile ion-disks yields 
\begin{equation} \label{15}
{\tilde u_e} = \frac{{\tilde n}_e}{n_o}  \frac{\omega}{ k}, \qquad {\tilde u_e} = \frac{ 4 k {\tilde n}_e}{M_e  \omega} \, Q^2 {\hat {V}}(kb) ,\end{equation}
which gives 
\begin{equation} \label{16}
\omega^2 = \omega_{Dpe}^2\, [(k b)^2 \,{\hat {V}}(kb)] ,\quad {\rm where} \quad \omega_{Dpe}^2= \frac {4 Q^2 n_o}{M_e b^2}.
\end{equation}
We see that the disk-plasma waves are dispersive for finite values of $kb$. In the short wavelength limit   $kb\gg1 $ the waves become Langmuir waves (see Note \ref{N1} and  Eq.(\ref{A7}),  Appendix \ref{App1}) with frequency $\omega_{Dpe}$ where $n_o/(\pi b^2)$ has the dimension of a volume density.
\\
In the long wavelength limit (see Appendix \ref{App1}, Eq.(\ref{A6})  which defines  the constants $C_1,\, C_2$) we find  (logarithmically corrected) ``cold electron-disk sound'' waves of the form 
\begin{equation} \label{17}
\omega^2 =  k^2[C_1 + C_2\ln{(kb)^2}]  v^2_D ,\quad {\rm where} \quad  v^2_D  = b^2 \omega_{Dpe}^2= \frac {4 Q^2 n_o}{M_e}.
\end{equation} 
We note that in a  particle plasma there is no  velocity directly corresponding to $v_D$. \\
The phase velocity of Langmuir waves is smaller than $v_D$  as follows from  $\omega^2/k^2 = \omega_{Dpe}^2/k^2 = v^2_D /(kb)^2<<v^2_D $
while  in the case of the cold sound waves\footnote{Note that  in the limit of  planar foils instead of disks, i.e. for $b\to \infty$ these sound waves  would not be present.}
we have $ \omega^2/k^2   \sim  [\ln(kb)]^2 v_D^2   >>   v_D^2$.

\section{Disk Vlasov equation and waves}

The treatment of the Vlasov kinetic equation follows standard lines with two minor differences: the expression of the interaction potential $V(\xi)$ is different from the Coulomb potential and it is more convenient to use the particle density  as the independent variable instead of the potential 
Thus we have the integro-differential equation 
\begin{equation} \label{18}
\frac{\partial { f}_j(x,v,t)}{\partial t} + v \frac{\partial {f}_j(x,v,t)}{\partial x} + \frac{{{\cal F}}_j(x,t)}{M_j} \frac{\partial { f}_{j}(x,v,t)}{\partial v} = 0
\end{equation}
where $ {f}_{j}(x,v,t)$ is the distribution function of the $j$ species ($j= e, i$), $v$ is the disk velocity  coordinate in phase space and   ${\cal F}_j(x,t) $ is defined in Eq.(\ref{7})]
and 
\begin{equation} \label{19}
 {\rho}(x,t)  =  Q \left[\int_{-\infty}^{+\infty}  { f}_{i}(x,v,t) \, dv -\int_{-\infty}^{+\infty}  { f}_{e}(x,v,t) \, dv \right] .
\end{equation}

\subsection{Linearized equation}

In a homogeneous disk plasma the linearized Vlasov  equation takes the form
\begin{equation} \label{18bis}
\frac{\partial {\tilde f}_j(x,v,t)}{\partial t} + v \frac{\partial {\tilde f}_j(x,v,t)}{\partial x} + \frac{{\tilde {\cal F}}_j(x,t)}{M_j} \frac{\partial { f}_{j0}(v)}{\partial v} = 0
\end{equation}
where $ { f}_{j0}(v)$ and ${\tilde f}_j(x,v,t)$ are the equilibrium and the perturbed distribution functions, $v$ is the disk velocity and 
\begin{equation} \label{19bis }
{\tilde {\cal F}}_j(x,t)  = - 4 i k  Q_j   {\hat \rho}(k)  {\hat  V}(kb), 
\end{equation}
is the perturbed force, as obtained from Eq.(\ref{9}).
\\
Considering again for the sake of simplicity immobile ion disks and waves of the form
\begin{align} \label{20}
& {\tilde f}_{e}(x,v,t) = {\tilde f}_{o}(v)  \exp {[-i (\omega t -  kx)]}, \qquad  {\tilde \rho}(x,t) =  {\tilde \rho}_o\, \exp {[-i (\omega t -  kx)]}
\\ & {\rm with} \quad   {\tilde \rho}_o = -Q   {\tilde n}_{eo} =  -Q \int_{-\infty}^{+\infty}  {\tilde f}_{eo}(v) \, dv,\nonumber 
 \end{align}
we obtain the dispersion relation
\begin{equation} \label{21}
 1 = - 4  k Q^2   \frac{{\hat {V}}(kb)}{M_e} \int_{-\infty}^{+\infty} \frac{ \partial {f}_{eo}(v)/\partial v}{\omega - k v} \, dv = -  \frac{\omega_{Dpe}^2}{k}  [(kb)^2{\hat {V}}(kb)] \int_{-\infty}^{+\infty} \frac{ \partial [{f}_{eo}(v)/n_o]/\partial v}{\omega - k v} \, dv 
, \end{equation}
where the velocity integral has to be taken according to the Landau prescription. In the cold limit  when $\omega/k \gg v_{the}$, with $v_{the}$
the halfwidth of the distribution function ${\tilde f}_{eo}$, Eq.(\ref{21})  becomes
 \begin{equation} \label{22}
 1 =  4   Q^2  n_o  \frac{k^2}{\omega^2}  \frac{{\hat {V}}(kb)}{M_e} = \frac{\omega_{Dpe}^2}{\omega^2}  [(kb)^2{\hat {V}}(kb)] ,\end{equation}
 which coincides with Eq.(\ref{16}).

The dispersion relation given by Eq.(\ref{21}) depends on  the normalized wave number $kb$ and in addition on   the equilibrium dimensionless parameter $(v_{the}/v_D)^2 = \Theta $ (see below Eq.(\ref{FF13a})). For disk-Langmuir  waves the cold limit $\omega/k\gg v_{the}$  implies $v_{the}/v_D \ll 1/(kb)\ll1$, whereas for cold sound waves it implies the less restrictive condition $v_{the}/v_D \ll |\ln{(kb)^2}|$ where  $|\ln{(kb)^2}| \gg 1$ .

Normalizing velocities over $v_D$,  Eq. (\ref{21}) can be rewritten  in terms of $ {\bar \omega} =\omega/\omega_{Dpe}$, \, ${\bar v} = v/v_d$ and $\kappa = kb$ as 
\begin{equation} \label{23}
 1 = - \kappa{\hat {V}}(\kappa) \int_{-\infty}^{+\infty} \frac{ \partial {\bar f}_{eo}({\bar v})/\partial {\bar v}}{{\bar \omega} - \kappa {\bar v}} \, {d\bar v}\ 
, \end{equation}
where  the dimensionless distribution function   
$
 {\bar f}_{eo}({\bar v}) = v_D f_{eo}(v)/n_o
$ depends on the  dimensionless  temperature \,  $\Theta =   T_e/(4 Q^2 n_o)$.

\section{Conclusions}\label{conc}

In order to model collective electrostatic effects  on a 3-D plasma  using a reduced dimensionality configuration, it may be convenient to define an effective  interaction energy constructed so as to retain the physical effects that we deem must be included in order to to construct a ``faithful''   model.  
\\
In these notes  we have introduced  a model of interacting charged disks  that provides a  one-dimensional model with an interaction energy that vanishes at infinity.

Clearly this approach  is restricted  to the case where the system  evolution may be described by  ``scalar one-dimensional electrodynamics'', i.e. when no  magnetic fields  are present and the electric field is the gradient  of a one-dimensional scalar potential.

\appendix 
\section{Disks electrostatics} \label{App1}
A set of relevant formulae  derived in  Ref.\cite{Disk} are listed   below along with some notational adaptations.

The axially symmetric electrostatic potential $\varphi(r,x)$ created by a uniformly charged disk  with its center at the coordinate origin, with  total charge $Q$ and radius  $b$ can be written as 
\begin{equation} \label{A1}
\varphi(r,x) =    2 Q    \int_{0}^{+\infty} ds  J_o(sr)\frac{J_1(sb)}{sb}\exp{(-s|x|)},
\end{equation}
where $J_0$ and $J_1$  are are Bessel functions, while the $x$-component of the electric field can be written as
\begin{equation} \label{A2}
{\epsilon}_x (r,x) =   {\rm sign (x)} \,  \frac{2 Q}{b}    \int_{0}^{+\infty} ds  J_o(sr) J_1(sb) \, \exp{(-s|x|)},
\end{equation}
which, at $r=0$, reduces to \begin{equation} \label{A3}
{\epsilon}_x (0,x) =   {\rm sign (x)} \,   \frac{2 Q}{b}    \int_{0}^{+\infty} ds  J_1(sb) \, \exp{(-s|x|)} =  {\rm sign (x)} \, \frac{2 Q}{b^2}  \left[ 1 - \frac{|x|/b}{(1 + x^2/b^2)^{1/2}}\right].
\end{equation}
The expression for the interaction energy in Eq.(\ref{1}) is the result of the integral
\begin{equation} \label{A4}
{\cal W} (|x|) =    4  \frac{Q_1\,  Q_2}{b} V(|x|/b) =  \frac{4 Q_1Q_2}{b}    \int_{0}^{+\infty} ds  [J_1(s)/s]^2 \, \exp{(-s|x|/b)} ,
\end{equation}
which is obtained by integrating Eq.(\ref{A1}) over the disk surface.

The Fourier transform of $V(\xi)$ can be  obtained from the equation above in the form
\begin{align} \label{A5} &
{\hat V} (kb) =   \int _{-\infty}^{+\infty} \frac{dx}{b} \,V(|x|/b) \exp{-[i (kb) \, (x/b)]}  =    \\ &    \int_{0}^{+\infty} ds  [J_1(s)/s]^2  \, \int _{-\infty}^{+\infty} \frac{dx}{b}  \, \exp{(-s|x|/b)} \exp{-[i (kb) \, (x/b)]}  \nonumber= \\&
2  \int_{0}^{+\infty} ds  [J_1(s)/s]^2  \,   \int _0^{+\infty} \frac{dx}{b}  \, \exp{(-s x /b)} \cos{[(kb) \, (x/b)]}  =\nonumber\\ &
2  \int_{0}^{+\infty} ds  [J_1(s)/s]^2  \,  \frac{s}{(kb)^2 + s^2} = \frac{1}{  \pi^{1/2} (kb)^2}G^{2,2}_{2,4}\left((kb)^2\left| ^{1/2,1}_{1,1,-1,0} \right.\right)
\nonumber
\end{align} 
 where Meijer $G^{2,2}_{2,4}$ is a Meijer G function   \cite{Nist}, denoted as  $G$-Function in the  notation  used by ``Mathematica''  (see \url{https://reference.wolfram.com/language/ref/MeijerG.html})\\

  $${\rm Meijer}G[
  \{\{a_1, a_2\},\{...\} \}, 
  \{\{b_1,…,b_2\},\{b_{3},b_{4}\}\}
  ,z] =  G^{2,2}_{2,4}\left(z\left| ^{a_1,a_2}_{b_1,b_2,b_3 , b_4} \right.\right)$$

The function $  G^{2,2}_{2,4}\left((kb)^2\left| ^{1/2,1}_{1,1,-1,0} \right.\right)/\left[ 2 (\pi^{1/2} (kb)^2) \right] $   has the following limits  (as given by ``Mathematica'' )
\begin{align} \label{A6} &
\frac{1}{  \pi^{1/2} (kb)^2} G^{2,2}_{2,4}\left((kb)^2\left| ^{1/2,1}_{1,1,-1,0} \right.\right) =    (1/8)\left[ 5 - 6 \gamma - 2 \ln{[(kb)^2]} - 2 \psi(3/2) \right] \nonumber
  \\ & + (1/48)\left[ 13 - 9 \gamma - 3 \ln{[(kb)^2]} - 3 \psi(5/2) \right]  (kb)^2 \nonumber  \\&
+(5/92316\left[ -65 +36  \gamma+12
\ln{[(kb)^2]}+12  \psi(7/2) \right]  (kb)^4 + ...,
\end{align} 
for $(kb)^2\ll 1$ ,
where $\gamma = 0.577216$, $\psi (z) = d \ln{\Gamma(z)}/dz$,  
and 
\begin{align} \label{A7} &
G^{2,2}_{2,4}\left((kb)^2\left| ^{1/2,1}_{1,1,-1,0} \right.\right) =   \pi^{1/2} + ....\end{align} 
for  $(kb)^2\gg 1$. 
In this  latter limit we recover the standard Coulomb-like dependence ${\hat V} \propto 1/k^2$. \\ The logarithmic dependence at small wavenumbers  arises from the $1/x$ dependence of the interaction energy  at large distances in 1-D.

\end{document}